\documentclass[pre,twocolumn,floatfix,superscriptaddress,aps]{revtex4}

\usepackage{amsmath}
\usepackage{makeidx}
\usepackage{amssymb}
\usepackage{graphicx}
\usepackage{bmpsize}
\usepackage{hyperref}
\usepackage{soul}
\usepackage[usenames]{color}
\usepackage[normalem]{ulem}

\newcommand{\beq}{\begin{equation}}
\newcommand{\eeq}{\end{equation}} 
\newcommand{\beqa}{\begin{eqnarray}}
\newcommand{\eeqa}{\end{eqnarray}} 

\begin{document}

\title{ Weak-coupling to unitarity crossover in Bose-Fermi mixtures: Mixing-demixing and spontaneous symmetry breaking in trapped 
systems}

\author{Sandeep Gautam \footnote{sandeep@iitrpr.ac.in}}
 
\affiliation{Department of Physics, Indian Institute of Technology Ropar, 
Rupnagar, Punjab 140001, India} 

\author{S.K. Adhikari \footnote{sk.adhikari@unesp.br; 
http://www.ift.unesp.br/users/adhikari}}
\affiliation{Instituto de F\'{\i}sica Te\'orica, UNESP - Universidade 
Estadual Paulista, 01.140-070 S\~ao Paulo, S\~ao Paulo, Brazil} 
 
\begin{abstract}
 The usual treatment of a Bose-Fermi 
mixture relies on weak-coupling Gross-Pitaevskii (GP) and density-functional 
(DF) Lagrangians, often including the more realistic {perturbative} Lee-Huang-Yang (LHY) 
corrections. We suggest analytic {non-perturbative} beyond-mean-field Bose and Fermi 
Lagrangians valid along the crossover from weak- to strong-coupling limits 
of intra-species interactions consistent with the LHY corrections and the 
strong-coupling (unitarity) limit for small and large scattering lengths 
$|a|$, respectively, and use these  to study the Bose-Fermi mixture. 
We study numerically mixing-demixing and spontaneous symmetry breaking in 
Bose-Fermi mixtures in spherically-symmetric and quasi-one-dimensional traps 
while the intra-species  Bose and Fermi interactions are varied from 
weak-coupling to strong-coupling limits.The LHY 
correction is appropriate for medium to weak atomic interactions and diverges
for stronger interactions (large scattering length $|a|$), whereas the 
present beyond-mean-field Lagrangian is finite in the unitarity limit
($|a|\to \infty$). We illustrate our results using the Bose-Fermi 
$^7$Li-$^6$Li mixture under a spherically-symmetric and a 
quasi-one-dimensional trap. The results  obtained with the present model for 
density distribution of the Bose-Fermi mixture along the crossover could be 
qualitatively different from the usual GP-DF Lagrangian with or without LHY 
corrections. Specifically, we identified spontaneous symmetry breaking and 
demixing in the present model not found in the usual model with the same 
values of the parameters.    
\end{abstract}


\maketitle

\section {Introduction} 
Soon after the observation of Bose-Einstein condensates (BEC) in ultra-cold, 
ultra-dilute, harmonically trapped alkali atom vapors \cite{BEC}, several 
groups were able to create and study trapped super-fluid Fermi gas 
\cite{FERMI} and Bose-Fermi mixture \cite{bfmixture} in a laboratory. Of 
these, a study of trapped super-fluid Bose-Fermi mixture is of interest 
because of a rich variety of phenomena it can exhibit \cite{rmpb,rmpf}. Such study
can provide information about different intra- and inter-species interactions 
acting in this mixture. Phase-separation $-$ a typical feature of binary 
super-fluids $-$ in mixtures of quantum degenerate gases has been investigated 
in Bose-Fermi systems \cite{Molmer,as}. With the advent of experimental 
techniques, now it is possible to change the different inter- and 
intra-species interactions in a super-fluid Bose-Fermi mixture by 
manipulating an external electromagnetic interaction near a Feshbach 
resonance \cite{fesh}. Hence, it is of natural interest to see how the 
different mixed and demixed phases of a super-fluid Bose-Fermi mixture change
as the inter- and intra-species interactions are varied. It is also of 
interest to see if such phases could spontaneously break the symmetry of the 
underlying Lagrangian. 

In the present paper, we study the mixing-demixing transition in Bose-Fermi
super-fluid mixtures in three-dimensional isotropic and quasi-one-dimensional
(quasi-1D) harmonic traps as the intra-species Bose and Fermi 
interactions are increased from weak to strong coupling. The hyperfine 
spin-1/2 Fermi super-fluid is considered to be in a fully paired 
(spin-up-down) state \cite{as,Barbut} rather than in a spin-polarized single 
hyperfine state \cite{Molmer,bfmixture}. We consider the intra-species 
interaction for Bose (Fermi) component as repulsive (attractive) 
which can be changed continuously from the weak-coupling limit to unitarity. 
The inter-species interaction between the Bose and the Fermi 
components is considered to be repulsive within the weakly coupling region. 
The strong-coupling unitary limit, where the gas parameter 
$x=|a|n^{1/3}\to +\infty$ with $a$ the $s$-wave scattering length, 
and $n$  the density, has recently drawn a great-deal of attention as it 
is characterized by universal laws arising from scale invariance. This limit 
is of great interest in different areas, such as, Bose and Fermi 
super-fluids \cite{univ,castin,salomon, salomon2,hulet,jinF}, 
superconductivity \cite{string}, string theory \cite{asoke}, neutron 
\cite{star} and Bose \cite{bstar} stars, and quark-gluon plasma \cite{quark}, 
and can be achieved in a laboratory \cite{fesh} in a Bose-Fermi super-fluid 
mixture.    
 
The usual mean-field treatment of super-fluid Bose-Fermi mixture \cite{as} is
confined to the weak-coupling limit of Bose and Fermi interactions 
described by the Gross-Pitaevskii (GP) \cite{gross,rmpb} Lagrangian for 
bosons and the density-functional (DF) \cite{rmpf} Lagrangian for fermions. As
the interaction strength is increased, we need a {non-perturbative }beyond-mean-field 
description. Lee, Huang and Yang (LHY) provided a {perturbative} 
Lagrangian for bosons \cite{LHYb} and fermions \cite{LHYf}. Although the LHY
Lagrangian is valid for slightly stronger interactions, it is not appropriate 
for very strong interactions in the unitarity limit, where it diverges.  
For this investigation, we proposed minimal analytic forms of Bose and 
Fermi Lagrangians valid from weak-coupling to unitarity with proper LHY 
\cite{LHYb,LHYf} and unitarity limits without fitting parameter(s). Most of 
previous suggestions for such crossover functions for both bosons 
\cite{bose,as} and fermions \cite{fermi,as} were numerical with fitting 
parameter(s) and hence did not have the analytic LHY limits. With the present 
analytic weak-coupling to unitarity crossover functions, we write the 
dynamical beyond-mean-field equations for the Bose-Fermi system, which we use
in this study of Bose-Fermi mixture in spherically-symmetric and quasi-1D 
traps. { For a quasi-1D confinement, we use this beyond-mean-field  
3D model rather than a strict 1D model  obtained from a quantum mechanical 
many-body 1D Hamiltonian. Such a strict 1D model has novel properties like 
fermionization of bosons \cite{TG}.  The nonlinearities of the strict 1D model 
are also different from the present 3D model. For a large finite transverse trap  
in the present quasi-1D case, as in experiments,  the present 3D model should be 
appropriate. How the present  results will approximate the results of the strict 1D
 model under infinitely strong transverse trap is an open question beyond the 
scope of the present study.}

 In the present study, we find that the density distribution of the 
Bose-Fermi mixture along the weak- to strong-coupling crossover could be
qualitatively different from that obtained employing  the usual GP-DF 
Lagrangian with or without the LHY corrections. For example, we found 
demixing in the Bose-Fermi mixture obtained using the present model where 
the GP-DF Lagrangian predicted mixing of the components. We also found 
spontaneous symmetry breaking in the present model not found in the GP-DF 
model.

The plan of the paper is as follows. In Sec. \ref{II}, we present the 
analytic expressions for energy and chemical potential of uniform Bose and 
Fermi super-fluids along the weak coupling to unitarity crossover and derive 
the nonlinear equations to study the mixing-demixing transition and 
spontaneous symmetry breaking in Bose-Fermi super-fluid mixtures. In Sec. 
\ref{III}, we present the numerical results along the weak- to 
strong-coupling crossover for the Bose-Fermi mixture under 
spherically-symmetric and quasi-1D traps. We compared our results with 
those obtained from the usual weak-coupling GP-DF Lagrangian for the
Bose-Fermi mixture and found that the present results could be qualitatively 
different. 
A summary of our findings is given in Sec. \ref{IV}.
 
\vskip .5 cm

\section{Analytical model along the weak to strong coupling crossover}
\label{II}
Bose and fully-paired Fermi super-fluids in the weak-coupling limit  
are well described by mean-field GP \cite{gross} and DF \cite{DF} \cite{rmpf} equations, respectively. These equations for bosons \cite{rmpb} and fermions 
\cite{as} are equivalent to the super-fluid hydrodynamic equations. The 
Bose and Fermi super-fluids are described by a macroscopic order 
parameter. In case of bosons, the order parameter is also the single-particle 
wave function in the Hartree approximation of the many-body dynamics. In case
of fermions, the macroscopic order parameter refers to a fully-paired 
bosonic entity known as Cooper pair \cite{lnc}. {Hence,} the 
macroscopic hydrodynamic description of a Fermi super-fluid is formulated
in terms of paired fermions in the form of Cooper pairs \cite{lnc} and not in
terms of single-particle Fermi wave function \cite{as}. Such a 
description has led to excellent results for many  collective \cite{xxx,collective} phenomena in many-fermion super-fluids, such as density \cite{density}
distribution or frequency of oscillation \cite{xxx}, where a many-body description becomes unmanageable.  
 
{  The macroscopic behavior of a Bose \cite{rmpb} or Fermi \cite{rmpf}  super-fluid is governed by the classical Landau \cite{LANDAU}
equations  of irrotational hydrodynamics  with the velocity field ${\bf v}_i=\hbar \nabla S_i/2m_i$, where 
$S_i$ is the phase of the order parameter   $\phi_i({{\bf r}, t})=\sqrt{n_i({\bf r},t)}e^{{\mbox i}S_i({\bf r}, t)}$  where   ${\mbox i} = \sqrt{-1}$,  $n_i$ is the density, $m_i$ is the mass of the fundamental entity responsible for super-fluidity: a bosonic atom or a Fermi  pair.  Here $i=B$ stands for bosons and $i=P$ for paired fermions. The continuity equation and the irrotational flow equation in this case, e. g., \cite{LANDAU}
\begin{align}
\frac{\partial n_i}{\partial t}&+\nabla \cdot (n_i{\bf v}_i)=0,\\
m_i\frac{\partial {\bf v}_i}{\partial t}&+\nabla \Big(V_i+\frac{1}{2}m_iv_i^2-\frac{\hbar^2\nabla^2 \sqrt{n_i}}{2m_i\sqrt{n_i}} + \mu_i \Big) &=0\label{EQ2},  
\end{align}
are entirely equivalent to the following dynamical equation for the order parameter $\phi_i({{\bf r}, t})$
\begin{align}\label{landau}
{\mbox i}\hbar \frac{\partial \phi_i}{\partial t}= \Big(-\frac{\hbar^2}{2m_i}\nabla^2+   V_i+ \mu_i \Big) \phi_i,
\end{align}
where $V$ is an external potential and $\mu _i,  $ is the bulk chemical potential for the  uniform Bose or Fermi gas.  The quantum pressure term $  -{\hbar^2\nabla^2 \sqrt{n_i}}/{2m_i\sqrt{n_i}} $ was not present
in Eq. (\ref{EQ2}) in the original classical flow equations but were introduced later for an accurate description of the dynamics. This term leads to the proper kinetic energy term in the Galilean invariant \cite{as} dynamical equation (\ref{landau}). For bosons Eq. (\ref{landau}) is the GP equation. For fermions  we can relate the fermion and pair variables by $m_P=2m_F, V_P=2V_F, \mu _P=2 \mu_F$ and Eq. (\ref{landau}) becomes the following equivalent density functional (DF) equation
\begin{align}\label{fermion}
{\mbox i}\hbar \frac{\partial \phi_F}{\partial t}= \Big(-\frac{\hbar^2}{8m_F}\nabla^2+   V_F+ \mu_F \Big) \phi_F.
\end{align}
}

In the    strong-coupling regime, the scattering length 
$a_i$ is much larger than all length scales ($|a_i|\to \infty$) and 
consequently, the system shows universal behavior \cite{univ,gior,castin} 
determined by the density $n_i$ independent of the parameter $a_i$. Here 
$i=B$ stands for  bosons and $i=F$ for paired fermions. By dimensional 
arguments, the bulk chemical potential of a uniform Bose or Fermi gas 
at unitarity is given by \cite{as,rmpf}
\begin{equation}\label{unit}
\lim_{|a_i|\to \infty}\mu_i(n_i,a_i)= \frac{\hbar^2}{m_i}
\eta_i n_i^{2/3},
\end{equation}
where $\eta_i$ is a universal parameter and $m_i$ the mass of an atom. 

\begin{figure}[!t]
\begin{center}
 \includegraphics[width=\linewidth,clip]{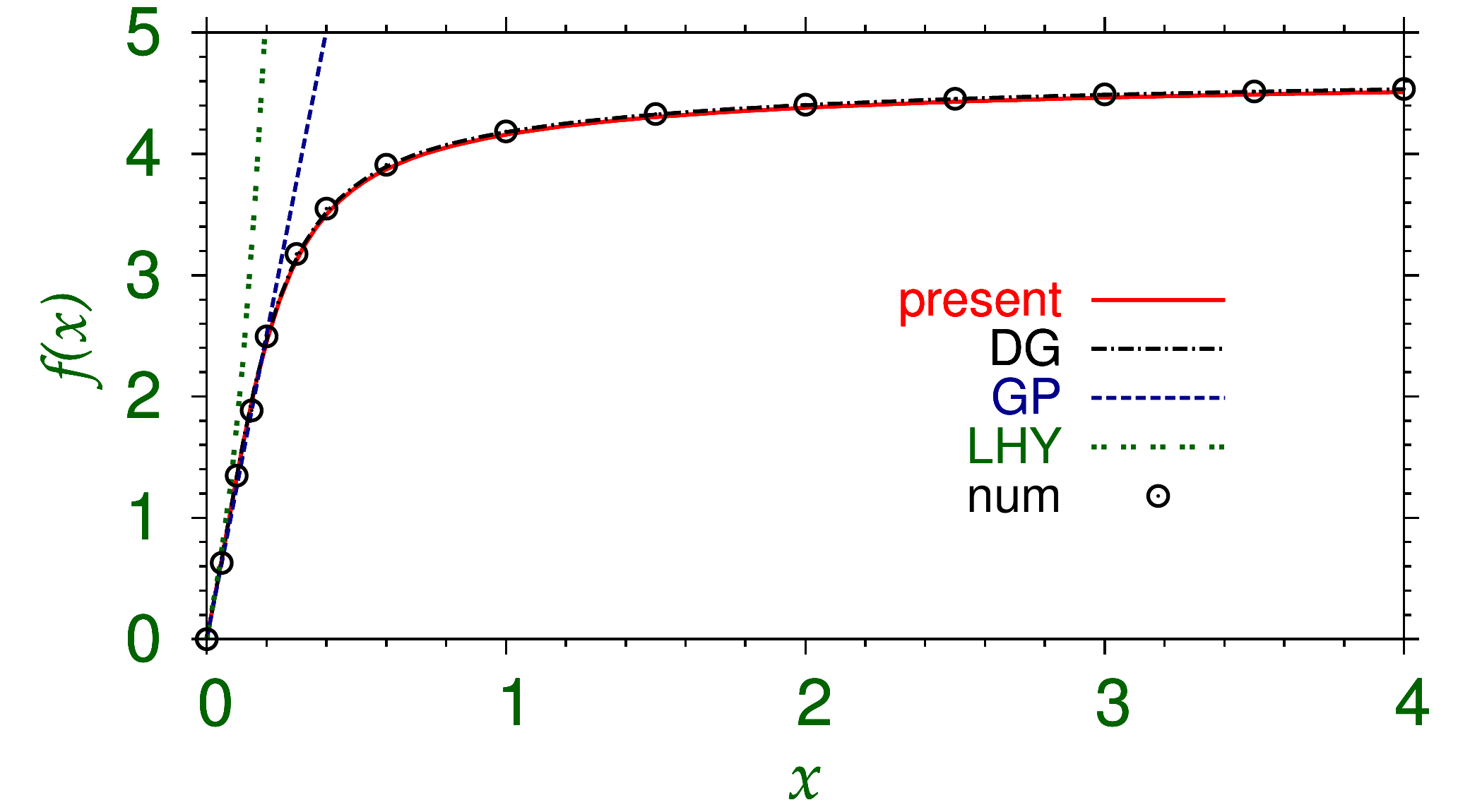}
\caption{(Color online) Dimensionless function $f(x)$ of the zero-temperature 
bulk chemical potential (\ref{muB}) versus $x$. 
present: the crossover function (\ref{coB}) with $\eta_B=4.7$,  DG: Hartree 
calculation of Ding and Greene \cite{DG}, GP: GP function $f(x)=4\pi x$, LHY: 
LHY function $f(x)=4\pi x+4\pi \alpha x^{5/2}$, num: numerically calculated 
from energy density (\ref{enB}) using $\mu_B=\partial (n_B {\cal E}_B)/\partial n_B$.}
\label{fig1} 
\end{center}
\end{figure}

In the weak-coupling limit  the bulk chemical potential  
of a uniform Bose gas is given by \cite{LHYb}
\begin{eqnarray}\label{LHYB}
\mu_B(n_B,a_B)=\frac{\hbar^2}{m_B} (4 \pi n_B a_B+2\pi \alpha n_B^{3/2}  
a_B^{5/2}+...),  
\end{eqnarray}
where $\alpha = 64/(3\sqrt \pi$), and the first term $ 4\pi n_B a_B$ on the 
right-hand side is the weak-coupling mean-field GP \cite{gross} result and 
the second term is the {perturbative} LHY contribution \cite{LHYb}, which 
becomes important for moderate values of the scattering length $a_B$ $(>0)$.
The  LHY contribution to $\mu_B(n_B,a_B)$ has limited validity as it diverges
as $a_B\to \infty$ at unitarity, whereas the correct $\mu_B(n_B,a_B)$ should 
remain finite at unitarity as given by Eq. (\ref{unit}). The {\it minimal} 
analytic form of the {non-perturbative} bulk chemical potential consistent with the LHY 
correction (\ref{LHYB}) and the unitarity limit (\ref{unit}) and also valid 
along the crossover from weak to strong coupling is 
\cite{scirep} 
\begin{eqnarray}\label{muB}
\mu_B(n_B,a_B)&\equiv&\frac{\hbar^2}{m_B}  n_B^{2/3}f(x),\quad  x=a_B n_B   ^ {1/3}, \\
f(x) &=&4\pi  \frac{x+\alpha x^{5/2}}{1+\frac{\alpha}{2}x^{3/2}
+\frac{4\pi \alpha}{\eta_B}x^{5/2}}. \label{coB}
\end{eqnarray}  
{{Equation (\ref{muB}) with (\ref{coB}) is a Pad{\'e} approximant to the bulk chemical potential 
with the proper weak-coupling LHY and unitarity limits. 
}}
The LHY bosonic energy density consistent with Eq. (\ref{LHYB}) 
is 
\begin{eqnarray}\label{LHYBE}
{\cal E}_B(n_B,a_B) &    \equiv & \frac{1}{n_B}\int_0^{n_B}  \mu _i(n,a_B) d n\nonumber \\
&=&
\frac{\hbar^2}{m_B} 
\left(2 \pi n_B a_B+\frac{4}{5}\pi \alpha n_B^{3/2}  a_B^{5/2}+...\right).  
\end{eqnarray}
The same at unitarity consistent with Eq. (\ref{unit}) is
\begin{eqnarray}\label{LHYBE2}
{\cal E}_B(n_B,a_B)= \frac{\hbar^2}{m_B} \frac{3}{5} \eta_B n_B^{2/3}.  
\end{eqnarray}
These two limiting values can be combined to give the following minimal energy
density valid from weak coupling to unitarity 
 \begin{eqnarray} \label{enB}
{\cal E}_B(n_B,a_B)&\equiv&\frac{\hbar^2}{m_B} \frac{2\pi n_B^{2/3}(x+\frac{4\alpha}{5} x^{5/2})}{1+\frac{2\alpha}{5}x^{3/2}
+\frac{8\pi \alpha}{3\eta_B}x^{5/2}} 
\end{eqnarray}  

Although there is no experimental estimate of the parameter $\eta_B$ 
for bosons despite some attempts \cite{univ}, there are several microscopic 
many-body calculations of this parameter lying in the range from 3 to 9 
\cite{otherB,DG}. Of these, Ding and Greene (DG) \cite{DG} performed a 
microscopic Hartree calculation along the crossover and in addition to the 
value of $\eta_B=4.7$, they provided a reliable estimate of the universal 
function $f(x)$ along the crossover. In Fig. \ref{fig1}, we illustrate the 
present universal function $f(x)$ of Eq. (\ref{coB}) for $\eta_B=4.7$ and 
compare with the same from the microscopic calculation of DG \cite{DG} and 
also with the GP functional $f(x)=4\pi x$ and the LHY functional 
$f(x)= 4\pi (x+\alpha x^ {5/2}/2)$. For very small $x$ or for small values 
$a_B$, both the GP and LHY functionals are in reasonable agreement with 
the present crossover functional as can be seen in Fig. \ref{fig1}. However,
for larger $x$, near unitarity, the GP and the LHY contribution cannot 
describe the actual state of affairs.

For a fully-paired uniform super-fluid of spin-1/2 Fermi gas, the energy 
density is given by \cite{rmpf,LHYf}
\begin{align}\label{LHYFE}
{\cal E}_F(n_F,a_F)&= \frac{3}{5}E_F\left[ 1+ c_1 y+c_2 y^2+ ...\right],\\
c_1&= \frac{10}{9\pi}, \quad c_2 = \frac{4(11-\mbox{ln} 4)}{21 \pi^2},
\quad y=k_Fa_F,
\end{align}
with Fermi momentum $k_F=(3 \pi^2 n_F)^{1/3}$, Fermi energy 
$E_F = \hbar^2 k_F^2 /2m_F$, $a_F$ $(<0)$ the scattering length of spin 
up-down fermions. In Eq. (\ref{LHYFE}), the first term $3E_F/5$ is the   
DF term \cite{as,rmpf} valid in the Bardeen-Cooper-Schrieffer (BCS) 
\cite{bcs} weak-coupling limit. The next two terms represent the 
{perturbative} LHY contribution \cite{LHYf}. The energy density at 
unitarity is written as \cite{xxx,as}
\begin{eqnarray}\label{unitFen}
\lim_{|a_F|\to \infty }{\cal E}_F(n_F,a_F)= 
                        \frac{3}{5}E_F \eta_F.
\end{eqnarray}
The minimal analytic energy density along the weak-coupling to unitarity 
crossover consistent with the weak-coupling LHY limit (\ref{LHYFE}) and the 
unitarity limit (\ref{unitFen}) is:
\begin{align}\label{enF}
{\cal E}_F(n_F,a_F)= 
\frac{3}{5}E_F\left[1+ \frac{c_1 y +(c_2-2c_1^2)y^2}{1-2c_1 y 
+\frac{(c_2-2c_1^2)y^2}{\eta_F-1}}     
 \right]. 
\end{align}

\begin{figure}[!t]

\begin{center}
 \includegraphics[width=.8\linewidth,clip]{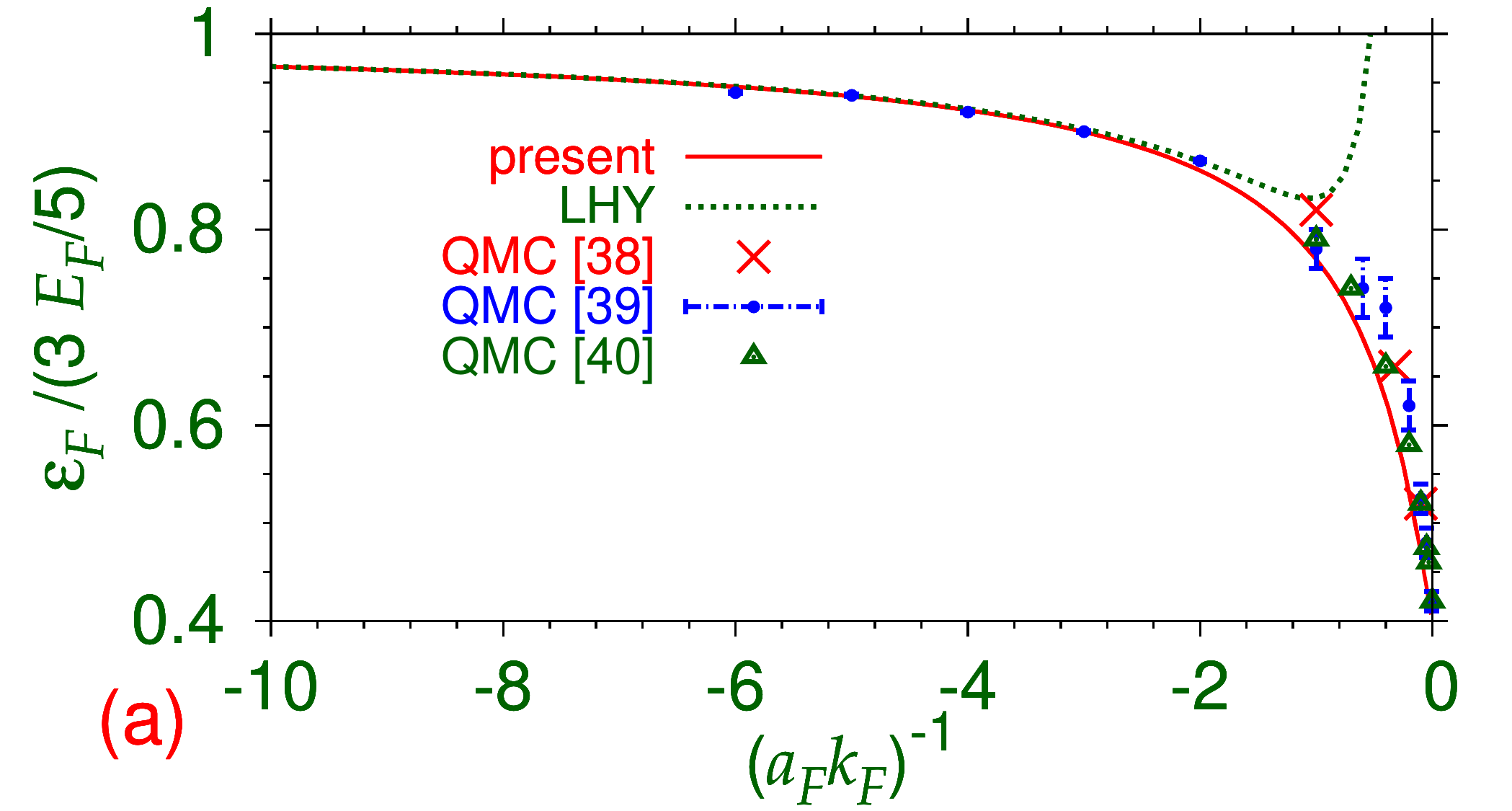}
 \includegraphics[width=.8\linewidth,clip]{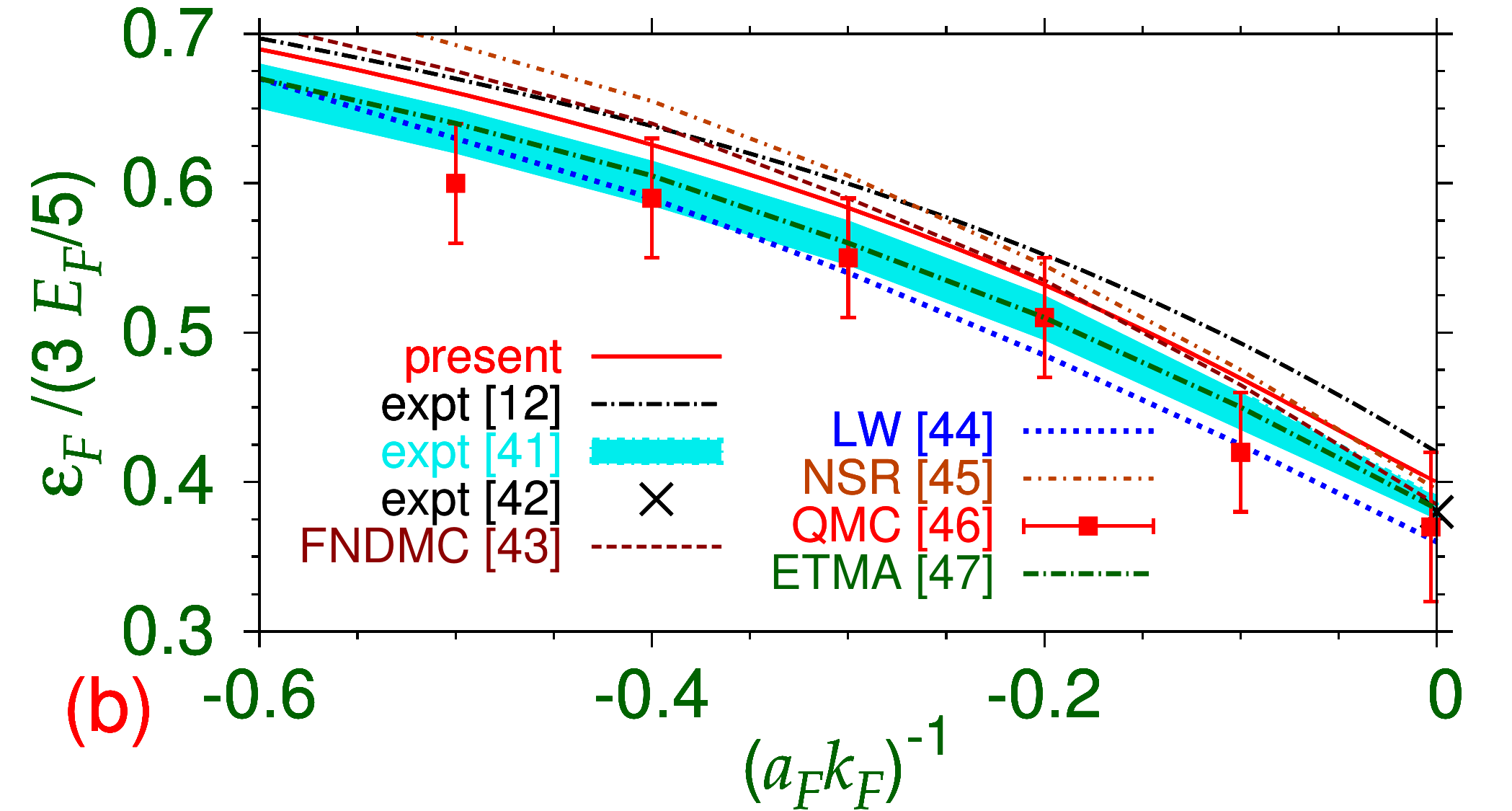} 
\includegraphics[width=.8\linewidth,clip]{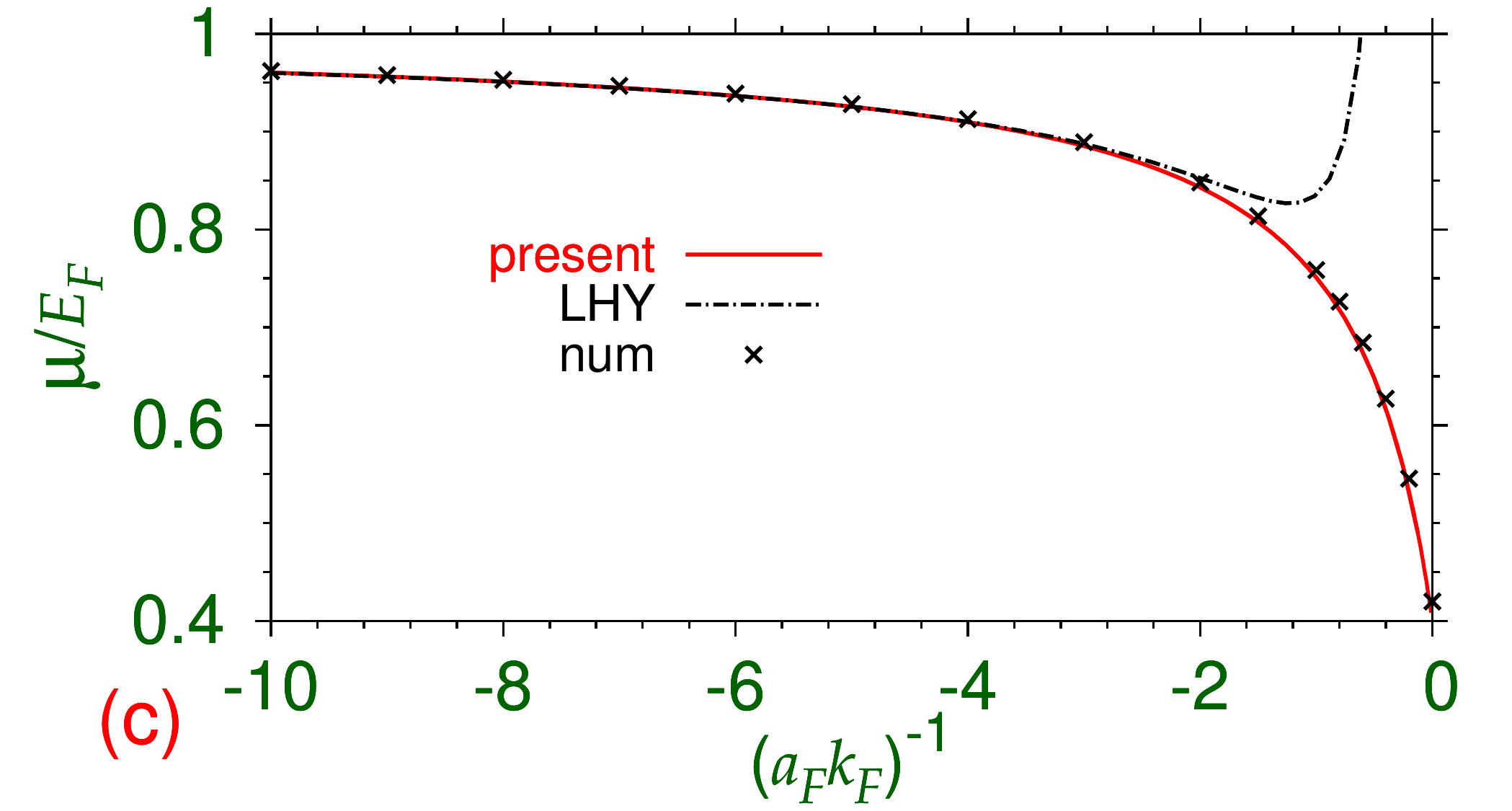}

\caption{(Color online) {Energy density ${\cal E}_F$ versus $(a_Fk_F)^{-1}$
 along the weak-coupling BCS to unitarity crossover: (a) Eq. (\ref{enF}) (present) with $\eta_F=0.4$;
Eq. (\ref{LHYFE}) (LHY); quantum Monte Carlo results of   Ref. \cite{microF2}; Ref. \cite{microF1}; and   Ref. \cite{microF3}. (b) Eq.  (\ref{enF}) (present) with $\eta_F=0.4$; experimental results of Refs. \cite{salomon},  \cite{prx}, and \cite{ku}, and theoretical results using FNDMC  \cite{FNDMC}, LW \cite{LW}, NSR \cite{NSR}, 
QMC  \cite{QMC}, and ETMA \cite{ETMA} methods.  (c)
{Bulk} chemical potential $\mu_F$ versus $(a_Fk_F)^{-1}$ along the weak-coupling BCS to unitarity crossover: Eq. (\ref{muF}) (present) with $\eta_F=0.4$; Eq. (\ref{LHYFMU}) 
(LHY); numerically calculated from Eq. (\ref{enF}) using $\mu_F= \partial (n_F{\cal E}_F)/\partial n_F$ (num).
 }}

\label{fig2}
\end{center}
\end{figure} 

The following  expression for the bulk chemical potential of a uniform Fermi 
gas in the weak-coupling LHY limit can be obtained from Eq. (\ref{LHYFE}):
\begin{align}\label{LHYFMU}
{\cal \mu}_F(n_F,a_F)&\equiv   \frac{\partial(n_F {\cal E}_F)}{\partial n_F}=
  E_F\left[ 1+d_1  y+d_2 y^2+ ...\right],\\
d_1&= \frac{4}{3\pi}, \quad d_2= \frac{4(11-\mbox{ln} 4)}{15 \pi^2}.
\end{align}
{ A non-perturbative} expression for the chemical potential along the crossover is written in 
an analogous fashion
\begin{align}\label{muF}
{\cal \mu}_F(n_F,a_F)\equiv & E_F g(y), \\
g(y)=&
\left[1+ \frac{d_1 y +(d_2-2d_1^2)y^2}{1-2d_1 y 
+\frac{(d_2-2d_1^2)y^2}{\eta_F-1}}\right].\label{coF}
\end{align}
 This bulk chemical potential has the correct LHY limit (\ref{LHYFMU}) and 
the unitarity limit:
\begin{eqnarray}\label{unitFmu}
\lim_{|a_F|\to \infty }{\mu}_F(n_F,a_F)=E_F.
\end{eqnarray}
{In this paper we will use Eq. (\ref{fermion}) to describe Fermi dynamics with  $\mu_F$ of Eq. (\ref{muF}) correct in the weak- and strong-coupling limits employing 
  Fermi  variables, and not pair variables. Actually, the chemical potential $\mu_F$ and the associated energy for fermions was used by von Weizs\"acker \cite{von} to describe properties 
of atomic nucleus   without any super-fluid properties  long before the work of Landau \cite{LANDAU}. The expression for chemical potential $\mu_F$   (\ref{muF}) remains valid for any value of super-fluid fraction. 
Hence,  Eq. (\ref{fermion}) can be used  to describe many macroscopic properties of equal mixture of spin-up and -down
 fermions, like density distribution and frequency of oscillation, independent of its super-fluid nature.   This is pertinent as for     
attractive interaction between spin-up and -down fermions the super-fluid fraction could be small \cite{2005}.
For small values of super-fluid fraction,  the use of   Eq. (\ref{fermion}) to describe super-fluid properties of attractive fermions, such as the generation of vortex lattice in a rotating Fermi super-fluid, may lead to qualitatively wrong result because the unpaired fermions remain in a normal state and do not contribute to vortex lattice formation. For describing the formation of vortex lattice,  a more fundamental set of dynamical equations \cite{2003} should be used. The  simple DF equation (\ref{fermion}) can, however, be used to describe non-super-fluid properties of the Fermi gas, such as density distribution and phase separation,  even for small values of super-fluid fraction. {There have been  
numerous successful applications of similar crossover models to study density distribution \cite{density,xxx} 
and  collective dynamics \cite{collective} of a Fermi gas along the crossover.}

{
There are results of energy density ${\cal E}_F(n_F,a_F)$ from several 
theoretical microscopic calculations  \cite{microF1,microF2,microF3,FNDMC,LW,NSR,QMC,ETMA} and  
experimental estimates \cite{salomon,prx,ku}. 
 Most of the theoretical and experimental 
estimates for $\eta_F$ lie in the range $\eta_F=0.4 \pm .05$ \cite{microF1,salomon,jinF,hulet,salomon2,prx,ku,FNDMC,LW,NSR,QMC,ETMA}
  and in this paper we will employ $\eta_F=0.4$.    In Fig. \ref{fig2}(a), we plot present energy density
${\cal E}_F$ of Eq. (\ref{enF}) versus $(k_Fa_F)^{-1}$, its LHY limit 
(\ref{LHYFE}) and the theoretical  \cite{microF1,microF2,microF3} 
  estimates of energy density.   In  \ref{fig2}(b)
we compare the present result with several other experimental \cite{salomon,prx,ku}  and theoretical 
\cite{FNDMC,LW,NSR,QMC,ETMA}
estimates near unitarity. }
In 
Fig. \ref{fig2}(c), we plot the chemical potential (\ref{muF}) versus 
$(k_Fa_F)^{-1}$, its LHY limit (\ref{LHYFE}), and the numerically calculated 
chemical potential from the energy density (\ref{enF}).
From Figs. \ref{fig2}, 
we find that, for large values of $(|a_F|k_F)^{-1}$ in the weak-coupling 
limit ($|a_F|\to 0$), the present crossover results for ${\cal E}_F$ and $\mu _F$   agree well 
with the LHY contribution. 
 Along the whole 
crossover,  the agreement of the present   ${\cal E}_F$ (\ref{enF}) and $\mu _F$
  (\ref{muF}) with other estimates is good, and we will use 
these in the present study with the crossover model.

There have been previous attempts to parameterize the chemical potential of a 
uniform Bose \cite{as,bose} and Fermi \cite{as,fermi} systems. These 
previous attempts heavily relied on fitting parameters and/or the agreement 
with known experimental and theoretical data was poor. The present analytic bulk 
chemical potentials for uniform bosons (\ref{muB}) and fermions (\ref{muF}) 
have no fitting parameters and have excellent agreement with known data as 
illustrated in Figs. \ref{fig1} and \ref{fig2}, and we will use these in this
study.   

  The Lagrangian density of the localized super-fluid Bose-Fermi  mixture is 
written as \cite{as}
\begin{align}
\label{lag}
{\cal L}&=
\sum_i  \Big[{\mbox i}\hbar \frac{N_i}{2}(\phi_i\dot \phi_i^*
- \phi_i^* \dot \phi_i)+ N_i\{ V_i+{\cal E}_i(n_i,a_i)\}|\phi_i|^2\Big]\nonumber \\
&+ \frac{N_B\hbar^2}{2m_B}|\nabla \phi_B|^2 
 +\frac{N_F\hbar^2}{8m_F} |\nabla \phi_F|^2
\nonumber \\ & 
+\frac{1}{2}
4\pi a_{BF}N_B N_F \frac{\hbar^2}{m_R}|\phi_B|^2   |\phi_F|^2 
  , 
\end{align}
where $\phi_i,N_i$ are the order parameter and number of atoms of the 
Bose or Fermi component, $m_R\equiv m_Bm_F/(m_B+m_F)$ is the reduced 
mass, $a_{BF}$ is the Bose-Fermi scattering length { {to characterize the spin-independent 
interaction between a
boson and a fermion},} 
 and the energies 
${{\cal E} _i}(n_i,a_i)$ are given by Eqs.  (\ref{enB})  and (\ref{enF})
and the density $n_i\equiv N_i|\phi_i|^2.$

The Euler-Lagrange equations for a spherically symmetric trap corresponding 
to Lagrangian (\ref{lag}) are
\begin{eqnarray}&  &\,
{\mbox i} \hbar \frac{\partial \phi_B({\bf r},t)}{\partial t}=
{\Big [}  -\frac{\hbar^2\nabla^2}{2 m_B}   +m_B V({\bf r})                             
+ \mu_B(n_B,a_B) 
 \nonumber\\ &  &\, %
+ \frac{2\hbar^2\pi a_{BF} N_F}{m_R}  \vert \phi_F \vert^2
{\Big ]}  \phi_B({\bf r},t),
\label{eq3a}\\
& &\,
{\mbox i}\hbar \frac{\partial \phi_F({\bf r},t)}{\partial t}={\Big [}  
- \frac{\hbar^2 \nabla^2}{8m_F}   +m_FV({\bf r})         +\mu_F(n_F,a_F)                 
 \nonumber \\ & &\,
+ \frac{2\hbar^2\pi  a_{BF} N_B}{m_R}  \vert \phi_B \vert^2
{\Big ]}  \phi_F({\bf r},t),
\label{eq4a}
\end{eqnarray} 
where $\mu_B$ and $\mu_F$ are given by Eqs. (\ref{muB}) and  (\ref{muF}), 
respectively, valid along the crossover, and the confining trap is taken as 
\begin{eqnarray}
V({\bf r})= \frac{1}{2}  \left[ \omega_x^2 (x^2+y^2)+\omega_z^2 z^2  \right] 
\end{eqnarray}
where for a spherically-symmetric confinement $\omega_x=\omega_z$ and  
for a quasi-1D  confinement  $\omega_x\gg \omega_z,$ with $\omega_x$ 
and $\omega_z$ the trapping frequencies along $x$ and $z$ axes, 
respectively.  
{ In the absence of the interaction between bosons and fermions ($a_{BF}=0$),  Eqs. (\ref{eq3a})
and (\ref{eq4a}) reduce to Eqs. (\ref{landau}) and (\ref{fermion}) for bosons and fermions, respectively.

In Eqs. (\ref{eq3a}) and (\ref{eq4a}) the Bose and Fermi chemical potentials $\mu_B$ and $\mu_F$ 
are valid along the crossover from weak to strong coupling, whereas for Bose-Fermi interaction 
we are using its value in the weak-coupling limit as in the GP equation, as there is no 
universally accepted form of the Bose-Fermi interaction in the weak and strong couplings. 
This is acceptable if  the  calculation  is limited to only the weak coupling limit of 
Bose-Fermi interaction.  From Fig. \ref{fig1},  we find that the GP functional agrees with 
the crossover formula for {chemical potential} for values of the gas parameter 
$x \lessapprox 0.2$ and   the present study   will be limited in this domain. 
 
}

We obtain a set of coupled dimensionless equations from Eqs. (\ref{eq3a}) 
and (\ref{eq4a}) by expressing length in units of 
$l_0\equiv \sqrt{\hbar/m_B \omega_x}$, time in units of $t_0=m_Bl_0^2/\hbar$,
$|\phi_i|^2$ in units of $l_0^{-3}$, and energy in units of 
$\hbar^2/m_Bl_0^2$, etc:
\begin{align}\;
{\mbox i}  \frac{\partial \phi_B({\bf r},t)}{\partial t}=&
{\Big [}  -\frac{\nabla^2}{2 }+{\cal V}({\bf r}) +n_B^{2/3}  f(a_Bn_B^{1/3})
 \nonumber\\ 
 & 
+ \frac{2\pi m_B a_{BF} n_F}{m_R}  
{\Big ]}  \phi_B({\bf r},t),
\label{eq3}\\
{\mbox i} \frac{\partial \phi_F({\bf r},t)}{\partial t}=&{\Big [}  
- \frac{m_B \nabla^2}{8m_F}+\frac{m_F}{m_B} {\cal V}({\bf r})  +\frac{m_Bk_F^2}{2m_F}     g(a_F k_F  )
 \nonumber \\  & 
+ \frac{2\pi m_B a_{BF} n_B}{m_R}  
{\Big ]}  \phi_F({\bf r},t),
\label{eq4}  \\
{\cal V}({\bf r}) &= \frac{1}{2}  \left[  (x^2+y^2)+  \frac{\omega_z^2}{\omega_x^2} z^2  \right] ,
\end{align} 
where $k_F =  (3\pi^2 N_F|\phi_F|^2)^{1/3}$, and functions $f$ and $g$ are 
given by Eqs. (\ref{coB}) and (\ref{coF}).

\section{Numerical Results}
\label{III}

Equations (\ref{eq3}) and (\ref{eq4}) do not have analytic solution and
different numerical methods, such as  split time-step Crank-Nicolson 
\cite{CN,CN2} and Fourier pseudo-spectral \cite{FS} methods, are usually used 
to obtain their solution. The ground state of the Bose-Fermi mixture is 
obtained by solving Eqs. (\ref{eq3})-(\ref{eq4}) in imaginary time \cite{CN}.

We consider  a Bose-Fermi   mixture of $^{7}$Li and $^{6}$Li  atoms in a 
  three-dimensional isotropic trap ($\omega_x=\omega_z$),  where we use the radial symmetry of the 
system, trapping potential and emergent solutions, to cast the Eqs. 
(\ref{eq3})-(\ref{eq4}) in terms of a single spatial coordinate $r$, i.e. the 
radial coordinate of the spherical polar coordinate system. { The radial 
spatial step size $\Delta r$ and time step $\Delta t$ used to solve the 
Eqs. (\ref{eq4})-(\ref{eq3}) numerically are 0.05  and 0.0001,
respectively,  in dimensionless units. }

\begin{figure}[!t]

\begin{center}
\includegraphics[ height=4cm, width=7.5cm]{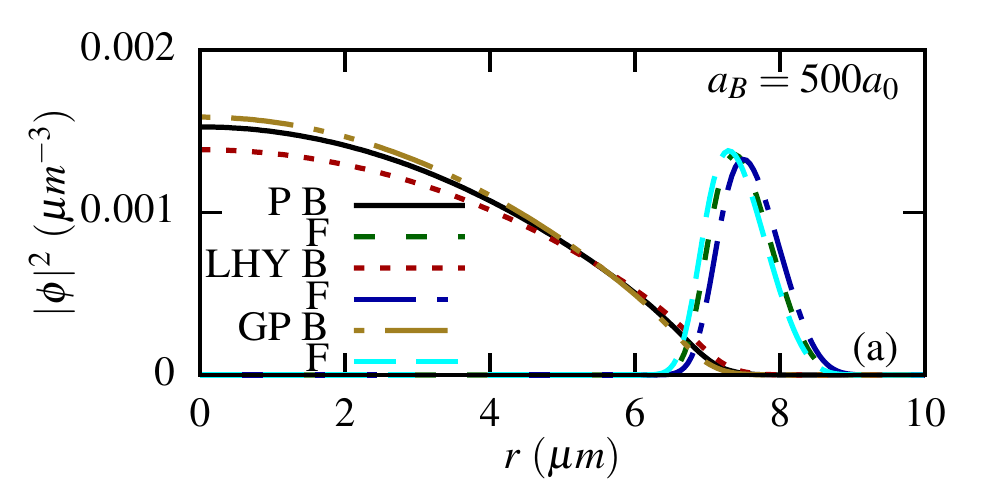}
\includegraphics[ height=4cm, width=7.5cm]{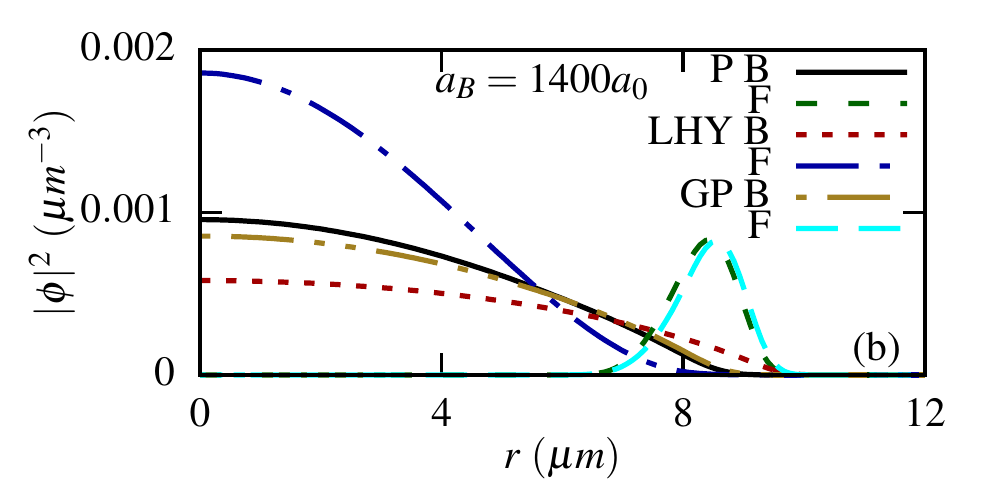}
\includegraphics[ height=4cm, width=7.5cm]{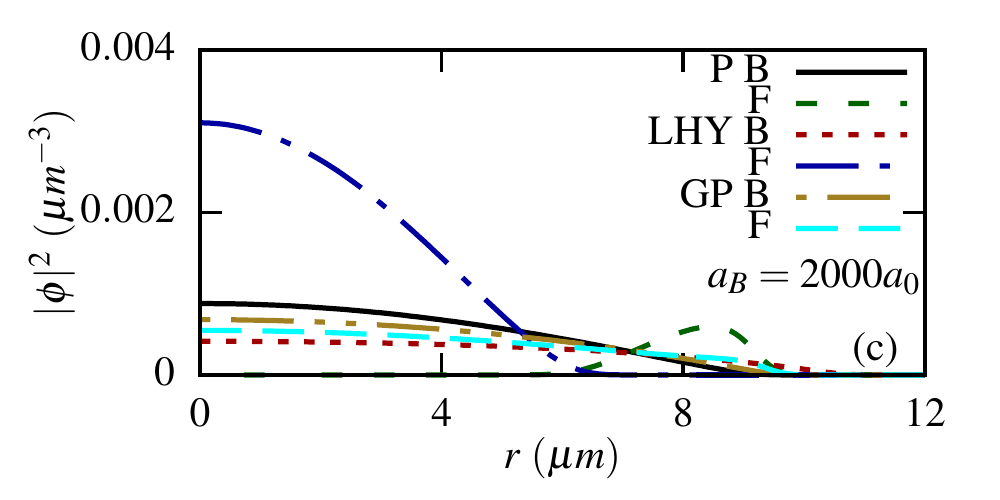}

\caption{(Color online) Densities of Bose ($|\phi_B|^2$) and Fermi
($|\phi_F|^2 $) components in a harmonically trapped  $^{7}$Li-$^{6}$Li mixture 
from a solution of    Eqs. (\ref{eq3})-(\ref{eq4}) by imaginary-time propagation:
 present (P) crossover  model, 
  Eqs. (\ref{coB}) and (\ref{coF});   LHY (LHY) model, 
Eq. (\ref{LHY}); and  GP-DF (GP) model,  Eq. (\ref{GP-DF});  for   $a_{\rm B}=$ (a) $500a_0$, 
(b) $1400a_0$,  and (c) $2000a_0$. The other parameters,
$N_B=50000$, $N_F = 1000a_0$, $a_{\rm F} = 0$, and $a_{\rm BF}=1500a_0$, are 
the same for (a)-(c). 
}
\label{fig3}
\end{center}
\end{figure}

\begin{figure}[!t]
\begin{center}
\includegraphics[ height=4cm, width=7.5cm]{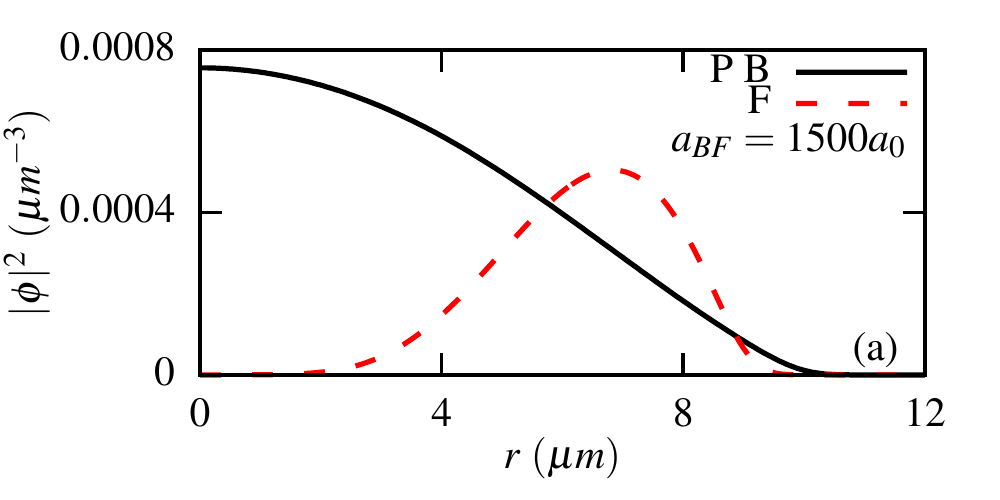}
\includegraphics[ height=4cm, width=7.5cm]{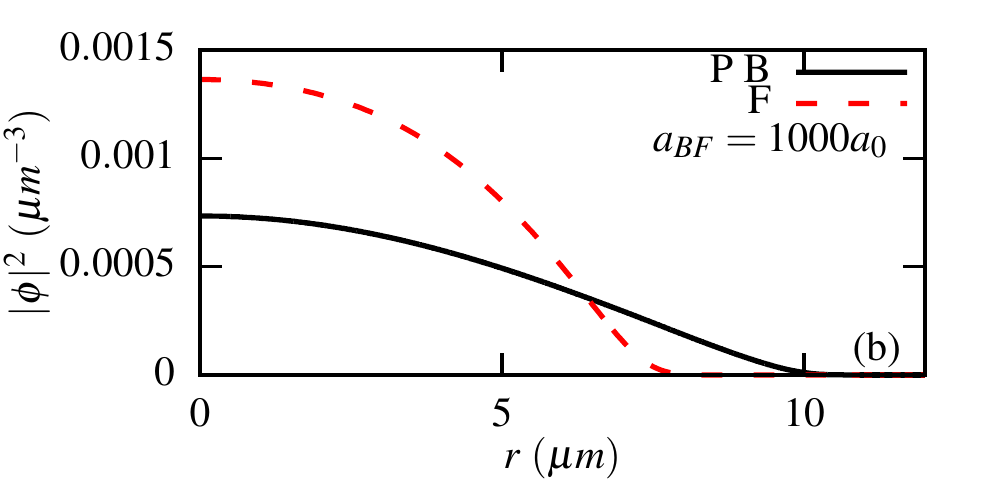}

\caption{(Color online) Same as in Fig. \ref{fig3} for parameters (a) $a_B=\infty, a_{BF}=1500a_0$
and (b)  $a_B=\infty, a_{BF}=1000a_0$ for the present (P) crossover model, other parameters are the same as in Fig. \ref{fig3}.
}
\label{fig4}
\end{center}
\end{figure}

The solutions obtained with the present crossover model which smoothly 
connects the weak-coupling regime with the unitarity regime are compared with 
the two models applicable in the weak-coupling regime, i.e. the GP-DF and LHY 
models. Before we proceed, let us precisely state what we mean by these three 
models. Solutions of Eqs. (\ref{eq3}) and (\ref{eq4}) with 
$f(x = a_{B} N_B^{1/3}|\phi_B|^{2/3})$ and $g(y = a_Fk_F)$ given, 
respectively, by Eqs. (\ref{coB}) and (\ref{coF}) are termed as solutions
obtained by the present model (denoted by symbol $P$ for present). In the 
LHY and GP-DF models, $f(x)$ and $g(y)$ in Eqs. (\ref{eq3}) and  (\ref{eq4}) 
are given, respectively, by
\begin{align}\label{LHY}
f(x) &= 4\pi\left(x + \frac{\alpha}{2} x^{5/2}\right),
\quad g(y) = 1 + d_1 y+d_2 y^2;\\
f(x) &= 4\pi x,\quad g(y) = 1.  \label{GP-DF}
\end{align}
We find that the ground state solution of the present model can be different 
quantitatively as well as qualitatively from the LHY and GP-DF models.

{{We find that without the inter-species interaction ($a_{BF}=0$) the density of both 
components is maximum at the center resulting in a mixed phase. With the increase of repulsive  inter-species interaction 
the density of one of the components may reduce at the center. With further increase of 
inter-species repulsion the density of one of the components  could be zero at the center and when that 
happens 
we will call the resultant Bose-Fermi state a demixed state. 
}}

\begin{figure}[!t]
\begin{center}
\includegraphics[ height=4cm, width=7.5cm]{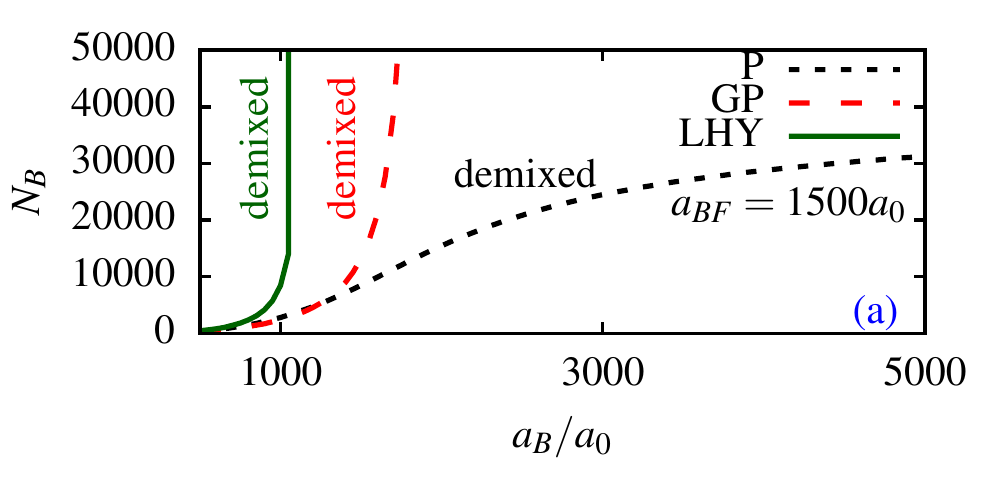}
\includegraphics[ height=4cm, width=7.5cm]{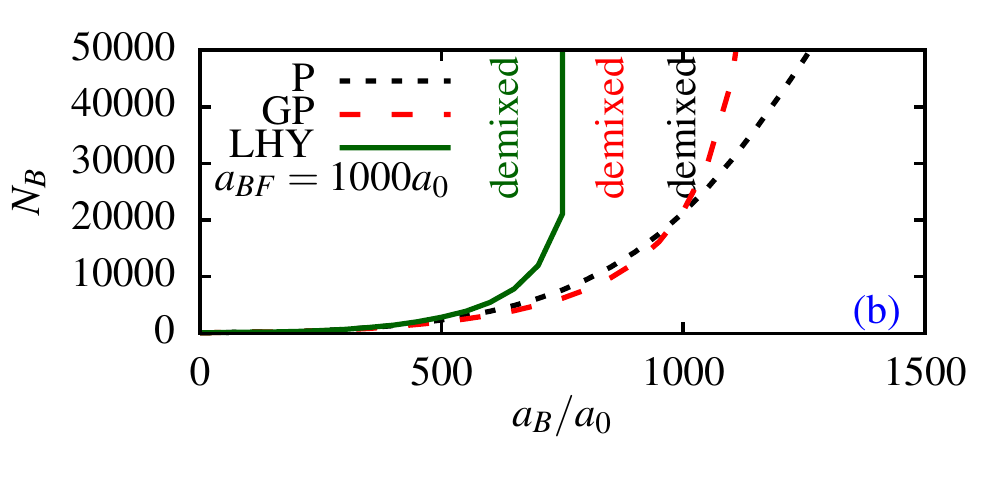} 
\caption{$N_{\rm B}$-$a_{\rm B}$ phase plots illustrating the mixing and 
demixing domains from the  present  (P) crossover, GP-DF (GP), LHY  (LHY) models for 
(a) $N_{\rm F} = 1000, a_{\rm F} = 0,$ and $a_{\rm BF} = 1500a_0$ 
 and (b) $N_{\rm F} = 1000, a_{\rm F} = 0,$ 
and $a_{\rm BF} = 1000a_0$.
The demixed  (mixed) states with the three models lie towards the left (right) of the respective lines. 
}
\label{fig5}
\end{center}
\end{figure}

\begin{figure}[!t]
\begin{center} 
\includegraphics[ height=4cm, width=8cm]{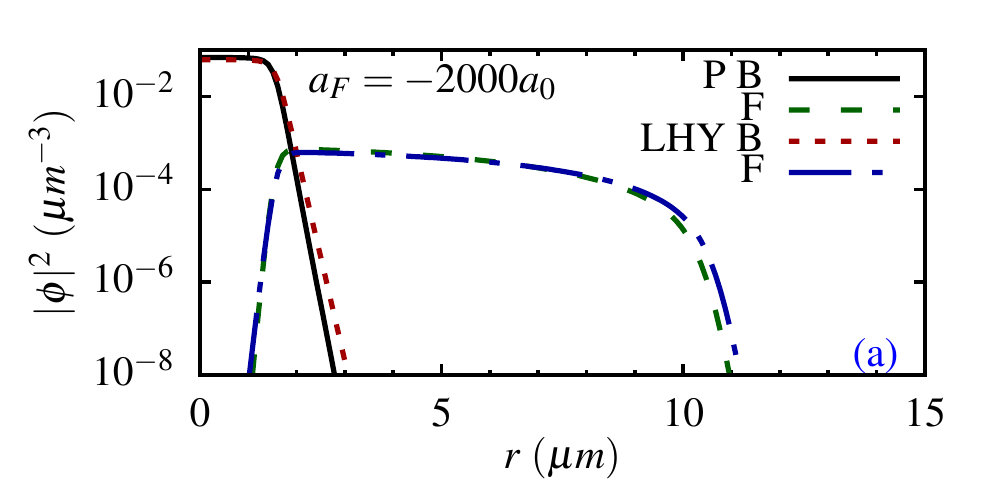}
\includegraphics[ height=4cm, width=8cm]{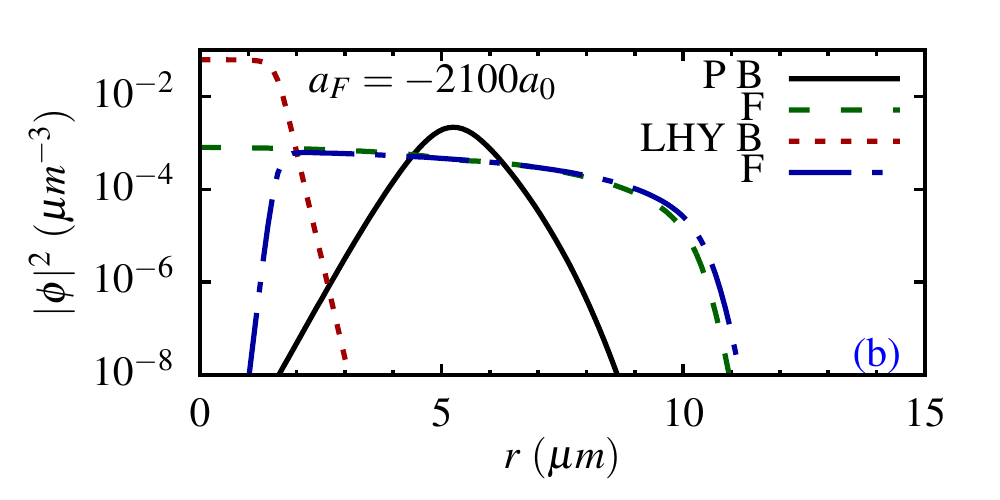}
\includegraphics[ height=4cm, width=8cm]{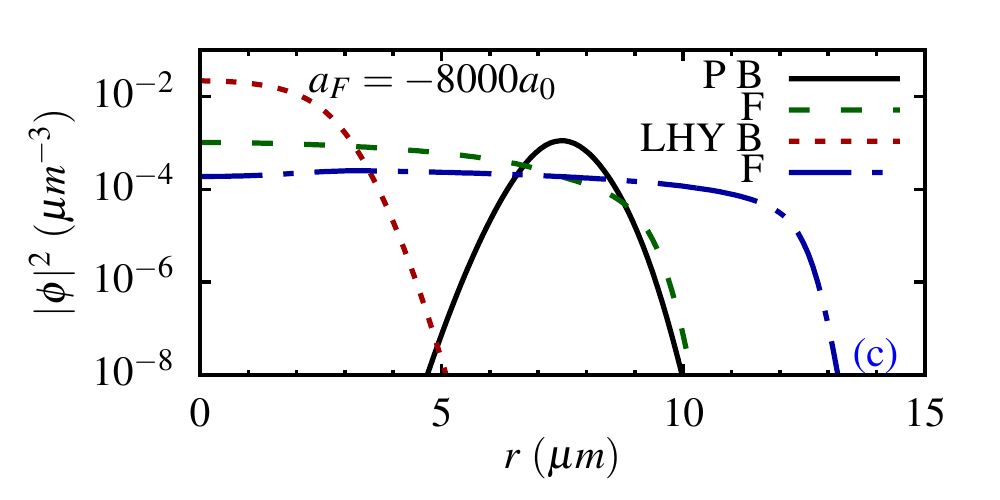}
\caption{(Color online)  Densities of Bose ($|\phi_B|^2$) and Fermi 
($|\phi_F|^2 $) components in a harmonically trapped  $^{7}$Li-$^{6}$Li mixture 
from the present (P) crossover model, 
   Eqs. (\ref{eq3})-(\ref{eq4}), 
and the LHY (LHY) model, 
Eq. (\ref{LHY}) for   $a_{\rm F}=$ (a) $-2000a_0$, 
(b) $-2100a_0$, and  (c) $-8000a_0$. The other parameters,
$N_B=1000$, $N_F = 50000a_0$, $a_{\rm B} = 500a_0$, and $a_{\rm BF}=1500a_0$, are 
the same for (a)-(c).
 }
\label{fig6}
\end{center}
\end{figure}

We first consider a Bose-Fermi $^7$Li-$^6$Li  mixture  with 
$N_{\rm B} = 50000, N_{\rm F} = 1000$, 
$a_{\rm F} = 0, a_{\rm BF} = 1500a_0$ for different  $a_{\rm B}$. In Figs. \ref{fig3} 
we plot component densities
$|\phi_i(r)|^2$ for the three models normalized as 
$4\pi\int r^2 dr  |\phi_i(r)|^2=1$. For $a_{\rm B} = 500a_0$, there is a 
good agreement between the component  densities obtained from the 
three models as is shown in Fig. \ref{fig3}(a), which indicates that the 
system is in the weak-coupling regime. As $a_{\rm B}$ is increased 
progressively to (b) $1400a_0$ and (c) $2000a_0$,  in both the LHY and 
GP-DF models, the system slowly changes from demixed state to mixed state 
with the transition first occurring for the LHY model, viz. Figs. \ref{fig3}
(a)-(b), and then for the GP-DF model, viz. Figs. \ref{fig3}(b)-(c); whereas 
in the present model the system remains always demixed.
 On further increase in 
$a_{\rm B}$, the system remains demixed as per the present model as shown in 
  Fig. \ref{fig4}(a) for 
$a_{\rm B}=\infty$. However, a large $a_{BF}$ ($=1500a_0$) is necessary for 
demixing   in the present model {  and} for a 
slightly smaller $a_{BF}$  ($=1000a_0$),  we have mixing in the present 
model, viz. Fig. \ref{fig4}(b). Hence the qualitative difference among the 
results of the present model on one hand and the LHY and the GP-DF models on 
the other hand as found in Figs. \ref{fig3}, as $a_B$ is changed from 
weak to strong coupling, is caused by a relatively large value of 
$a_{BF}$ ($=1500a_0$) used in numerical simulation. If we used the value 
$a_{BF}=1000a_0$ instead, keeping all other parameters unchanged, we verified 
that there will not be any qualitative difference in the results of the three
models: as $a_B$ is increased from weak to strong coupling in all three 
models there will be transition from demixed to mixed configuration 
(figure not presented). { The same will be true for the physical value 
$a_{BF}=40a_0$ for the Bose-Fermi scattering length in the $^7$Li-$^6$Li system \cite{expbf}.}

{ {Here we are using a large value of $a_{BF} $ in the range $1000a_0 \sim 1500a_0$, and
are using a mean-field GP-type Bose-Fermi  repulsion valid  for  small values of the gas parameter $x_{BF} \equiv  a_{BF}(n_{BF})^{1/3} \lessapprox 0.2$, where $n_{BF}$ is the geometric mean of Bose and Fermi densities. From Figs. \ref{fig3}  and \ref{fig4}, we find that the typical average  values of $|\phi_i|^2$ are less than 0.0005    and  the densities are $n_i = N_i  |\phi_i|^2$  with  $N_B=50000$, $N_F =1000$ and $a_{BF}=1500a_0$; consequently,  $x_{BF} \approx 0.12 < 0.2 $, where the GP approximation for Bose-Fermi interaction is valid.   In these figures the bosonic gas parameter with $N_B=50000$ and $a_B=1500a_0$ is larger than 0.2, thus requiring the present crossover formula for a proper description of dynamics.     }  
}

\begin{figure}[!t]
\begin{center}
\includegraphics[ height=4cm, width=8cm]{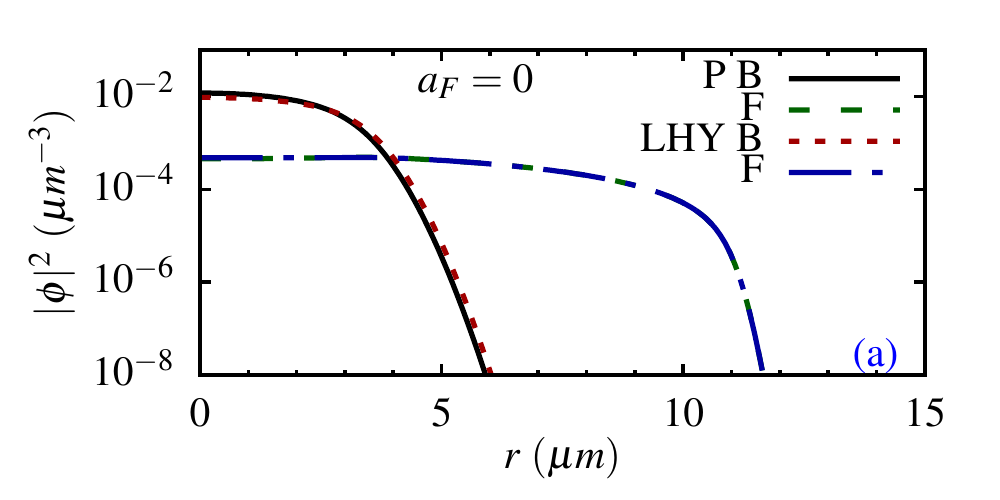} 
\includegraphics[ height=4cm, width=8cm]{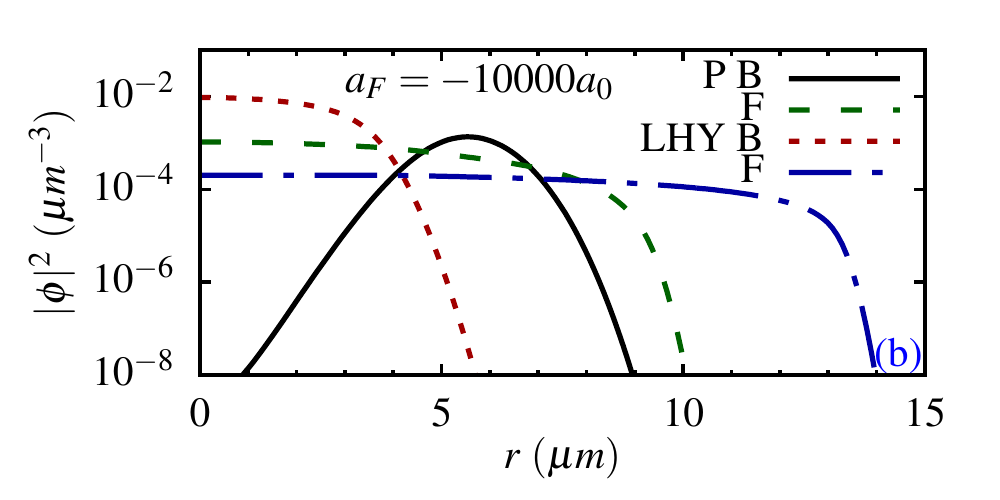}
\caption{(Color online) Same as in Fig. \ref{fig6} for $N_B=1000, N_F=50000, a_{BF}=1000a_0,a_B=1000a_0,$ and (a)
$ a_F =0$, and (b)  $a_F=-10000a_0$.
}
\label{fig7}
\end{center}
\end{figure}

\begin{figure}[!h]
\begin{center}
\includegraphics[ height=4cm, width=8cm]{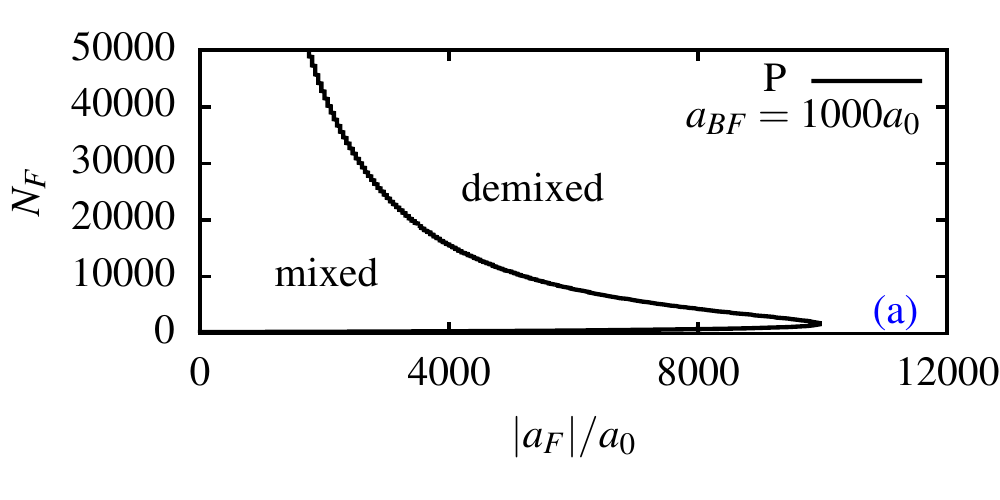}
\includegraphics[ height=4cm, width=8cm]{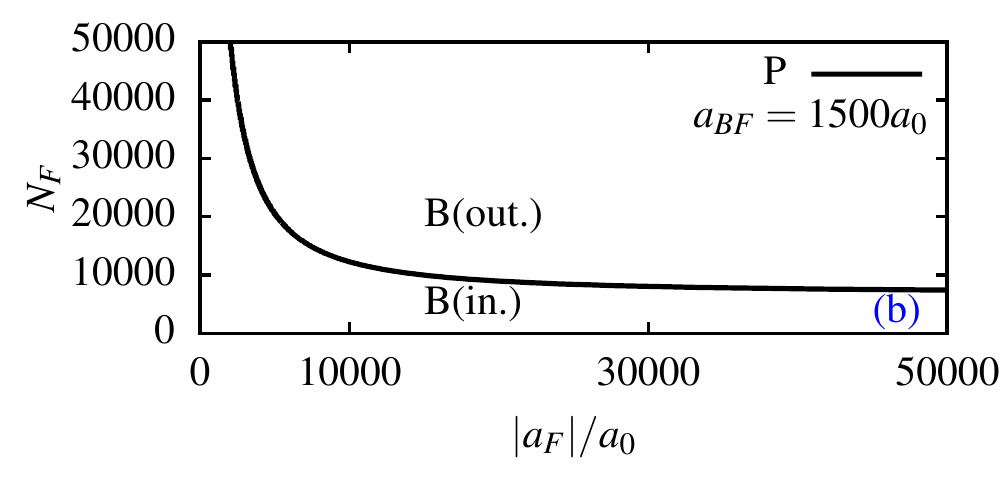}
\includegraphics[ height=4cm, width=8cm]{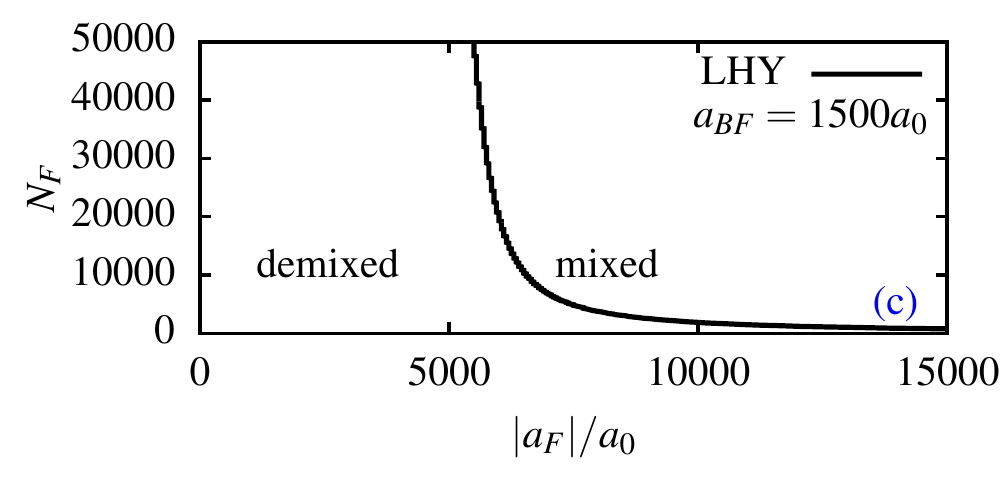}
\caption{{$N_{\rm F}$-$|a_{\rm F}|$ phase plots for $N_B=1000, a_B=500a_0$
and illustrating (a) the mixing and demixing domains for the present (P)  model
with $a_{BF}=1000a_0$, (b) crossover between the demixed state with bosons (fermions)
forming the core (shell) to the demixed state with bosons (fermions) forming the
shell (core) for present model with $a_{BF}=1500a_0$, and (c) the demixing and mixing 
domains for the LHY  (LHY) model with $a_{BF}=1500a_0$.}  
}
\label{fig8}
\end{center}
\end{figure}
 
 Keeping $N_{\rm F}$ and $ a_{\rm F}$  fixed, the $N_{\rm B}$-$a_{\rm B}$ 
phase plots showing the parameter domains of mixed and demixed states are 
illustrated in Figs. \ref{fig5}(a) and (b) for $a_{BF}=1500a_0$ and 
$1000a_0$, respectively. The demixed (mixed) states for the present, GP-DF, 
and LHY models appear on the left (right) side of the respective lines.

\begin{figure}[!h]
\begin{center}
\includegraphics[ height=4cm, width=8cm]{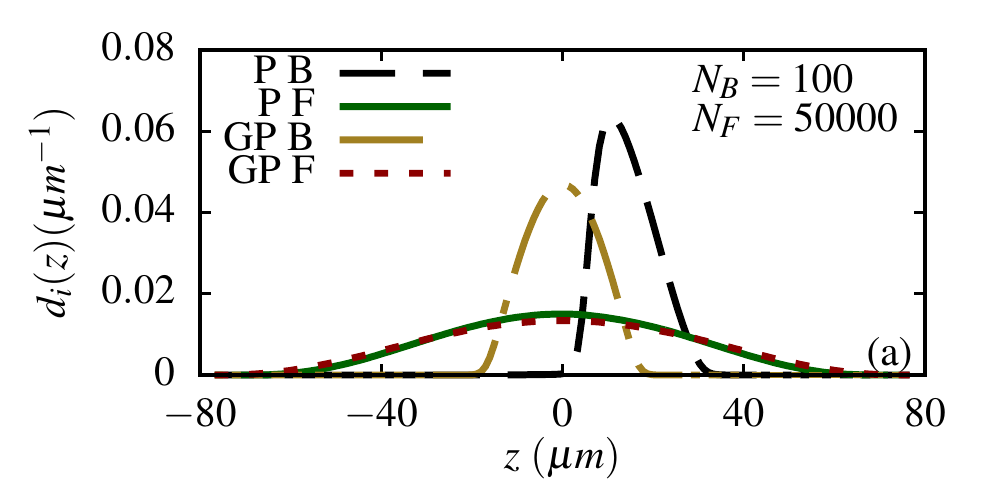}
\includegraphics[ height=4cm, width=8cm]{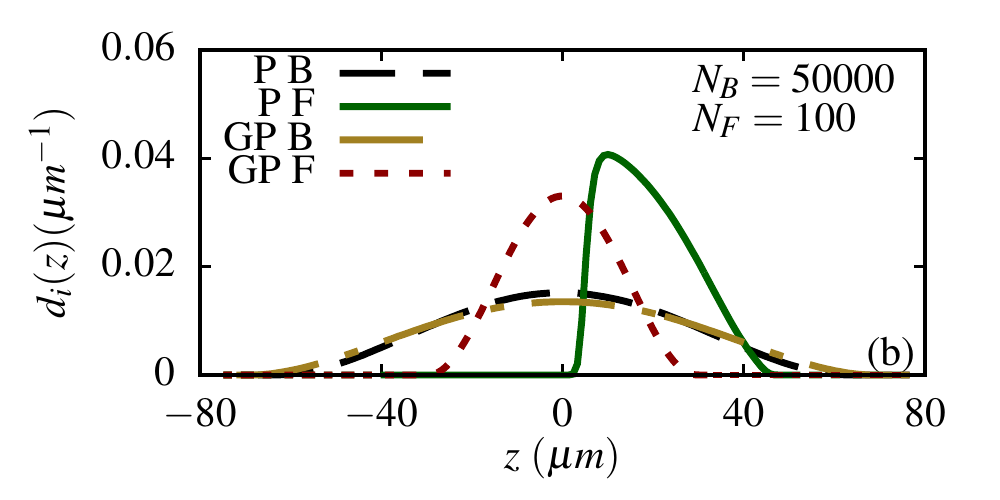} 
\caption{ Reduced density $d_{i}(z)$ of the binary Bose-Fermi mixture for (a) 
$N_B = 100$, $N_F = 50000$, $a_{\rm B}=4000a_0, a_{\rm F}=-4000a_0, 
a_{\rm BF} = 1500a_0$, (b) $N_B = 50000$, $N_F = 100$, $a_{\rm B}=4000a_0, 
a_{\rm F}=-4000a_0, a_{\rm BF} = 1500a_0$ for the present (P) and GP-DF (GP) models.
{ Here $\omega_z = 0.1\omega_x$}}
\label{fig9}
\end{center}
\end{figure}

Let us consider another case with $N_{\rm B} = 1000, N_{\rm F} = 50000$, 
$a_{\rm B} = 500a_0, a_{\rm BF} = 1500a_0$, while $a_{\rm F}$ is 
progressively decreased. In this case, for  {$-2000a_0 \le a_F <0$}, the 
three models lead to qualitatively similar demixing with the Bose 
component forming a core with the  Fermi component forming a shell 
surrounding it as shown in {Fig. \ref{fig6}(a)} for $a_F=-2000a_0$. 
The results of the GP-DF model are independent of $a_{\rm F}$, and hence
in Fig. \ref{fig6} we do not show the results of this model.
As $a_{\rm F}$ is decreased slightly to $a_F=-2100a_0$, 
we find a  Bose component forming a shell outside a Fermi core at the 
center as per the present model as is shown in Fig. \ref{fig6}(b) for 
$a_F=-2100a_0$. Upon further reduction of $a_{\rm F}$ to $-8000a_0$, the 
Bose shell moves further away from the center of the trap according to 
the present model, whereas LHY model ends up in the mixed phase as is shown 
in {Fig. \ref{fig6}(c).}

If we perform the analysis illustrated in Fig. \ref{fig6} for 
$a_{BF}=1000a_0$, the inter-species repulsion is weaker and in the 
weak-coupling limit ($a_F=0$), we have overlapping states for all models as 
shown in Fig. \ref{fig7}(a). Again we do not show the results of the GP-DF 
model as the same do not change with $a_F$. In the strong-coupling limit 
($a_F=-10000a_0$), {the present model results in the demixed state, 
whereas the LHY model remains in the mixed state, viz. Fig. \ref{fig7}(b).}

The mixing-demixing phenomena illustrated in Figs. \ref{fig6} and \ref{fig7}
leads to the $N_F$-$|a_F|$ phase plots of Figs. \ref{fig8}. In Fig.
\ref{fig8}(a) the fixed parameters are $N_B=1000, a_B=500a_0$ and 
$a_{BF}=1000a_0$. In this case mixing-demixing transition takes place only 
for the present model for $N_F \ge 270.$ 
{In Fig. \ref{fig8}(b), with fixed parameters 
$N_B=1000, a_B=500a_0, a_{BF}=1500a_0$, the present model, which results in 
a demixed ground state, predicts a position-swapping transition with the 
Bose component forming the core which is surrounded by the fermionic shell 
below a critical $N_F$ for a given value of $a_F$ ({as is shown in Fig. \ref{fig6}(a)}).  Above this critical 
$N_F$, the Fermi component moves to the core with the Bose component forming 
the shell around it as in Fig. \ref{fig6}(c). These two qualitatively different states are marked
as B(in) and B(out), respectively, in Fig. \ref{fig8}(b). For
the same fixed parameters  $N_B=1000, a_B=500a_0, a_{BF}=1500a_0$, the 
LHY model leads to the demixing-mixing transition as one increases $N_F$
at the fixed value of $a_F$ as is shown in Fig. \ref{fig8}(c).}

 Next we consider a quasi-1D $^{7}$Li-$^{6}$Li 
Bose-Fermi mixture \cite{Q1D} along the $z$ axis with strong traps in the $x-y$ plane ($\omega_x= 10 \omega_z $) 
and consider a few illustrative examples to show the 
qualitatively different ground state solutions obtained from the present 
model as compared to the  GP-DF and LHY models. 
{ The spatial step size 
$\Delta x = \Delta y = \Delta z$ and time step $\Delta t$ used to solve Eqs.  (\ref{eq3})-(\ref{eq4}) numerically are  $0.15$ and $0.005625$, 
respectively, in dimensionless units.} In such a quasi-1D Bose-Fermi mixture, the essential collective
dynamics and mixing-demixing transition take place in the $z$ direction. 
Hence for quasi-1D $^{7}$Li-$^{6}$Li Bose-Fermi mixtures, we will 
{illustrate} only the reduced 1D density along the $z$ direction 
$d_{i}(z) =\int dx dy |\phi_i(x,y,z)|^2$. Apart from mixing-demixing 
transition in the $z$ direction, we also find spontaneous symmetry-broken 
states in the quasi-1D $^{7}$Li-$^{6}$Li Bose-Fermi mixtures. Two examples 
of spontaneous symmetry breaking in the present model are shown in Fig. 
\ref{fig9}(a)-(b) for the reduced 1D density along $z$ direction where one 
of the components moved away from the center breaking the parity symmetry.
In this case the results of the GP-DF and LHY models lie very close 
to each other and the result of only the former model is shown.
In  Figs. \ref{fig9} there is {a} demixing in the present crossover model with 
the density of one of the components being zero at the center while the other component 
having a density maximum at the center. 
The densities of the LHY   model remain overlapping and parity 
symmetric.

In addition to the spontaneous symmetry broken states of Fig. \ref{fig9},
it is also possible to have partially to fully demixed states in quasi-1D 
Bose-Fermi mixtures. {For this purpose, we consider 
(a) $\omega_x= 10\omega_z$ and (b) $\omega_x= 100\omega_z$
, where the spatial step size $\Delta x = \Delta y = \Delta z$ and time step 
$\Delta t$ used to solve the equations (\ref{eq3})-(\ref{eq4}) numerically 
are  taken to be $0.1$ and $0.0025$, respectively. For both (a) and (b), we 
consider $N_{\rm B} = 1000$, $N_{\rm F} =100, a_B=500a_0, 
a_{\rm F} = -20000a_0$, and $a_{\rm BF} = 1500a_0$. The reduced 1D density 
$d_{i}(z)$ obtained in these two cases with the present and GP-DF models are shown in 
Figs. \ref{fig10}(a) and (b); it is evident that the bosons stay in the 
central region and the fermions are expelled symmetrically in two directions.
With the present model, the ground state is partially demixed with 
$\omega_x= 10\omega_z$, viz. Fig. \ref{fig10}(a) and fully demixed with 
$\omega_x= 100\omega_z$, viz. Fig. \ref{fig10}(b). In the GP-DF  
models, there is partial demixing which becomes more pronounced as the trap 
becomes more confined along the radial direction; { the same is the case with
LHY model (not shown in the figure).}}

\begin{figure}[!t]
\begin{center}
\includegraphics[ height=4cm, width=8cm]{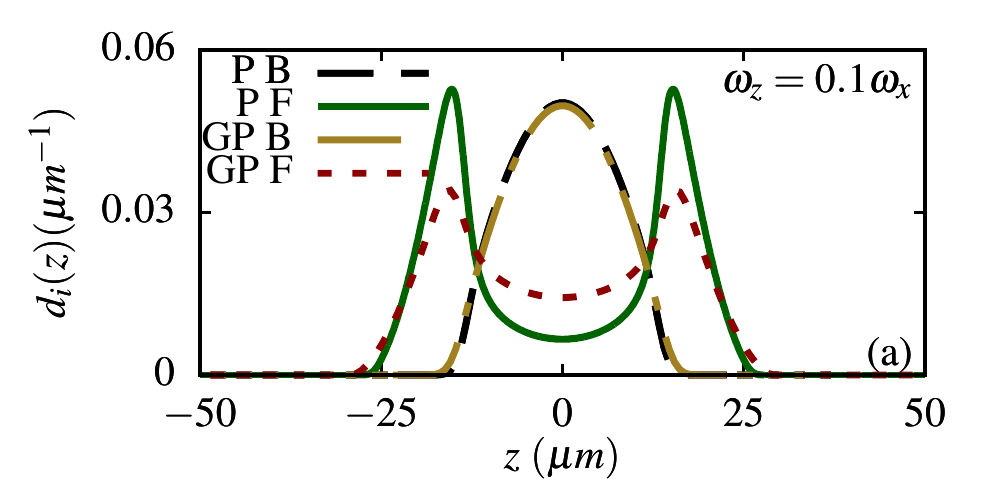}
\includegraphics[height=4cm, width=8cm,clip]{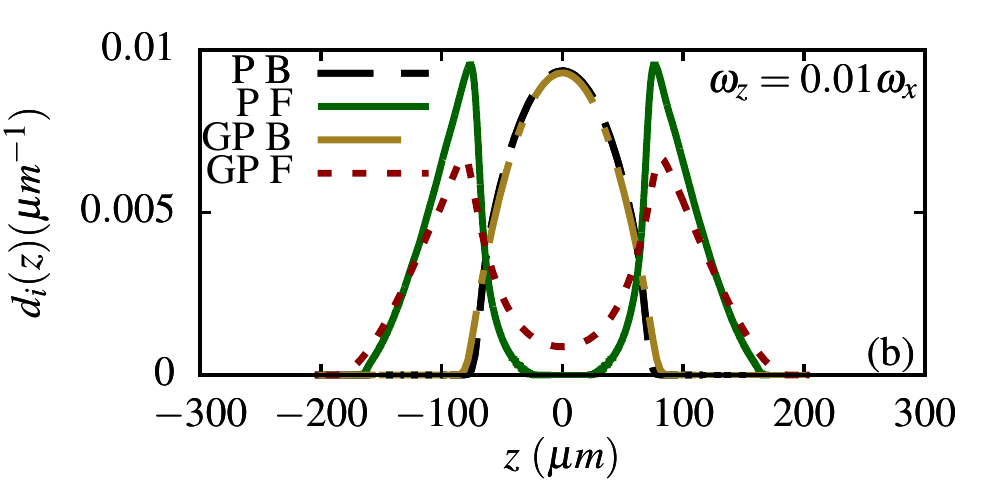} 
\caption{ {Reduced density $d_{i}(z)$ of the binary mixture for $N_B = 1000$,
$N_F = 100$, $a_{\rm B}=500a_0, a_{\rm F}=-20000a_0, a_{\rm BF} = 1500a_0$
with (a) $\omega_z = 0.1\omega_x$ and (b)  $\omega_z = 0.01\omega_x$ for the present (P) 
and GP-DF (GP)  models.}}
\label{fig10}
\end{center}
\end{figure}

\section{SUMMARY}
\label{IV}

Here we have demonstrated spontaneous symmetry breaking and mixing-demixing 
transition in trapped super-fluid Bose-Fermi mixture along the crossover from 
weak-coupling to unitarity for both intra-species  Bose and Fermi 
interactions. For Bose-Fermi inter-species interaction, we have used the 
weak-coupling GP interaction. The usual description of the super-fluid 
Bose-Fermi mixture employs the weak-coupling GP Lagrangian for the bosons 
and DF Lagrangian for the fermions. The interaction term in the GP 
Lagrangian is essentially the same as the many-body Hartree interaction 
term;  that in the DF Lagrangian is the total kinetic energy of the fermions 
in the Fermi sea. Usual treatment of the Bose-Fermi mixture employing GP 
Lagrangian for bosons and DF Lagrangian for fermions is termed GP-DF 
formulation. To study the Bose-Fermi mixture along the weak-to-strong coupling 
crossover, we suggested analytic {non-perturbative} Lagrangians for intra-species  Bose and 
Fermi interactions with {correct} LHY limit(s) in the 
weak-coupling domain and with correct unitarity {limit(s)} in the strong-coupling 
domain. These analytic Lagrangians have a single parameter: the universal 
parameter(s) $\eta_i$ at unitarity for bosons and fermions. In this study we have 
used the most precise value(s) of this universal parameter, which should be updated 
in future applications, if possible.
The Euler-Lagrange equations of the Bose-Fermi mixture describe the 
dynamics and have been used in this study. 

In this study, we considered spherically-symmetric and quasi-1D traps. In both 
cases, we compared the results of the present model valid along the crossover 
with those of the usual weak-coupling GP-DF and LHY models and found that the 
two types of treatments may lead to qualitatively different results. For example, 
in spherically symmetric traps we identified cases of demixing in Bose-Fermi 
mixture using the present crossover model not found in the weak-coupling GP-DF 
and  LHY models. In quasi-1D traps, we found spontaneous symmetry breaking in the 
present crossover model not found in GP-DF and LHY models. Hence we conclude that 
the {perturbative}  LHY model is unable to properly describe the Bose-Fermi 
mixture away from the weak-coupling limit.

\begin{acknowledgments} 
SG thanks the Science \& Engineering Research Board, 
Department of Science and Technology, Government of India (Project: 
ECR/2017/001436) and ISIRD  of Indian Institute of Technology, Ropar (Project: 9-256/2016/IITRPR/823) for support. 
SKA thanks the Funda\c c\~ao de Amparo \`a Pesquisa do Estado de 
S\~ao Paulo (Brazil) (Project:  2016/01343-7) and the Conselho Nacional 
de Desenvolvimento   Cient\'ifico e Tecnol\'ogico (Brazil) 
(Project: 303280/2014-0) for support.  
\end{acknowledgments}

\end{document}